\documentclass[apj,iop]{emulateapj}

\usepackage{natbib}
\usepackage{longtable}
\usepackage{hyperref}

\usepackage[normalem]{ulem}
\usepackage[export]{adjustbox}
\usepackage{graphicx}
\usepackage{amsmath}
\usepackage{amssymb}
\usepackage{float}
\usepackage{xcolor}
\usepackage{booktabs}

\shorttitle{eFEDS AGN Candidates in Dwarf Galaxies}
\shortauthors{Sanchez et al.}

\begin{document}

\title{A Deeper Look into eFEDS AGN Candidates in Dwarf Galaxies with Chandra} 

\author{Adonis A. Sanchez}
\affiliation{eXtreme Gravity Institute, Department of Physics, Montana State University, Bozeman, MT 59717, USA}

\author{Amy E. Reines}
\affiliation{eXtreme Gravity Institute, Department of Physics, Montana State University, Bozeman, MT 59717, USA}

\author{\'Akos Bogd\'an}
\affiliation{Center for Astrophysics $\vert$ Harvard and Smithsonian, 60 Garden Street, Cambridge, MA 20138, USA}

\and

\author{Ralph P. Kraft}
\affiliation{Center for Astrophysics $\vert$ Harvard and Smithsonian, 60 Garden Street, Cambridge, MA 20138, USA}

\begin{abstract}

The ability to accurately discern active massive black holes (BHs) in nearby dwarf galaxies is paramount to understanding the origins and processes of ``seed" BHs in the early Universe. We present {\it Chandra X-ray Observatory} observations of a sample of three local dwarf galaxies (M$_{*}$ $\leqslant$ 3 × 10$^{9}$ M$_\odot$, z $\leqslant$ 0.15) previously identified as candidates for hosting active galactic nuclei (AGN). The galaxies were selected from the NASA-Sloan Atlas (NSA) with spatially coincident X-ray detections in the eROSITA Final Equatorial Depth Survey (eFEDS). Our new {\it Chandra} data reveal three X-ray point sources in two of the target galaxies with luminosities between log(L$_{\rm \text{2-10 keV}}$ [erg s$^{-1}$]) = 39.1 and 40.4. Our results support the presence of an AGN in these two galaxies and a ULX in one of them. For the AGNs, we estimate BH masses of $M_{\rm BH} \sim 10^{5-6} M_\odot$ and Eddington ratios on the order of $\sim 10^{-3}$.

\end{abstract}


\section{Introduction} \label{sec:intro}
It is well-understood and documented that massive black holes (BHs) with $10^6 \lesssim M_{\rm BH}/M_\odot \lesssim 10^9$ inhabit the centers of nearly all massive galaxies. A portion of these BHs exhibit accretion, emit light across the electromagnetic spectrum, and are luminescent as active galactic nuclei (AGNs). As we reach toward lower mass scales ($M_{\rm BH} \lesssim 10^{5} M_{\odot}$), the fraction of dwarf galaxies with stellar masses $M_\star \lesssim 10^{9} M_{\odot}$ hosting massive BHs/AGNs becomes much less certain. However, these systems are paramount as they may provide meaningful constraints on the seeding model(s) for the formation of the first generation of BHs in the early Universe \citep[e.g.,][]{2010A&ARv..18..279V,2020ARA&A..58..257G,2022NatAs...6...26R}. They also are representative of an early stage in the growth of massive BHs and provide an avenue for studying the effects of BH feedback at low masses \citep{2017ApJ...839L..13S}.

There has been consistent growth in the evidence supporting massive BHs in dwarf galaxies found via various selection techniques, such as optical emission line spectroscopy, 
optical variability, mid-IR colors, and radio observations \citep[for a review, see][]{2022NatAs...6...26R}. However, each of these methods has its limitations. Optical searches tend to be biased toward more massive BHs with higher Eddington fractions and galaxies with low star formation rates \citep{2013ApJ...775..116R}. Selection based on mid-IR colors suffers from contamination by dwarf starburst galaxies mimicking AGN signatures \citep{2016ApJ...832..119H,2021ApJ...914..133L}. Finally, the sensitivities of current radio surveys may miss a significant fraction of the population \citep{2020ApJ...888...36R}.

X-ray observations can bypass many issues suffered by other techniques, as they are effective at detecting active BHs with low Eddington ratios \citep{2008ApJ...680..154G, 2008ARA&A..46..475H, 2009ApJ...696..891H} or those in galaxies with active star formation \citep{2016ApJ...830L..35R, 2017ApJ...835..223S, 2021ApJ...912...89K}. X-ray studies have been conducted in the past not only to search for AGNs in dwarf galaxies but also to explore the BH occupation fraction and place AGN candidates in dwarf galaxies under further scrutiny to confirm their presence. For example, \citet{2015ApJ...805...12L} found candidate AGNs in local dwarf galaxies ($z < 0.055; M_{\star} \lesssim 10^{9.5} M_{\odot}$) from archival data within the Chandra Source Catalog, and a similar study was conducted by \citet{2020MNRAS.492.2268B} with the 3XMM catalog. Follow-up observations by \citet{2023MNRAS.519.5848T} aimed to verify the nature of the AGNs in three dwarfs from the sample found by \citet{2015ApJ...805...12L} using high-resolution \textit{Chandra} and \textit{HST} observations.

\begin{figure*}[!t]
\begin{center}
$\begin{array}{ccc}
\includegraphics[width=5in]{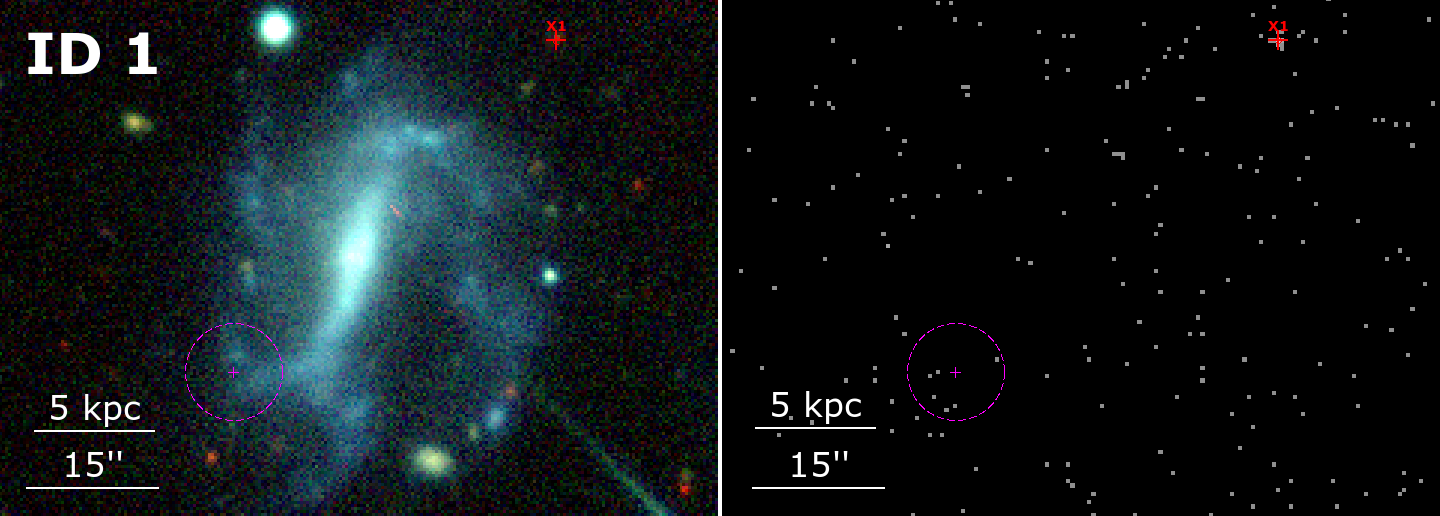} \\
\includegraphics[width=5in]{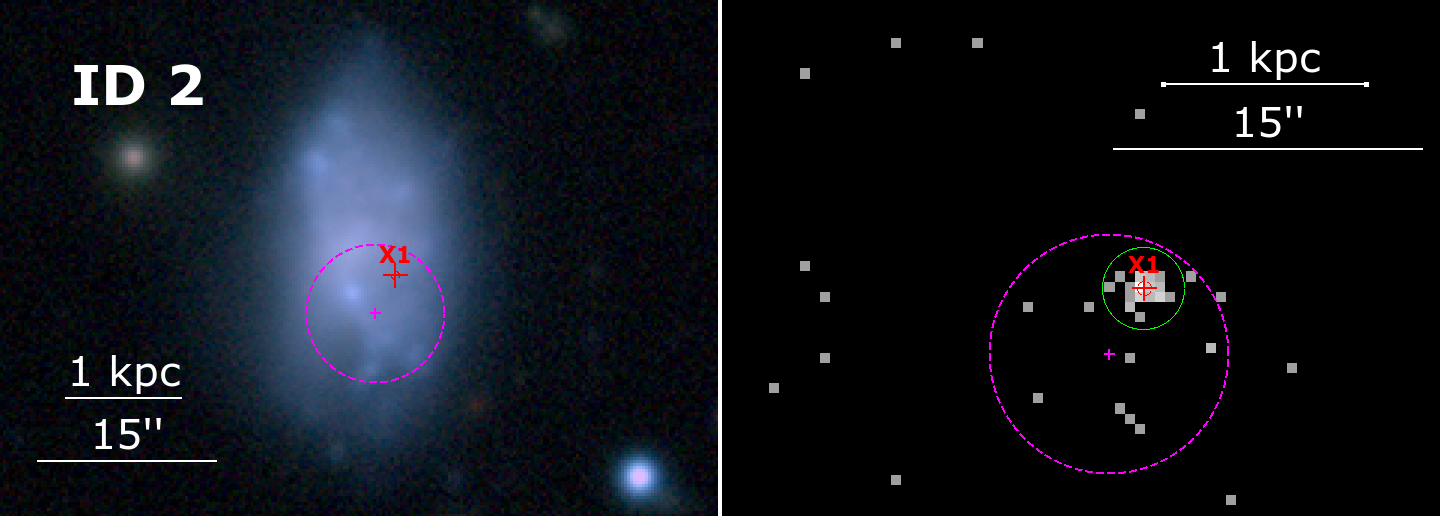} \\
\includegraphics[width=5in]{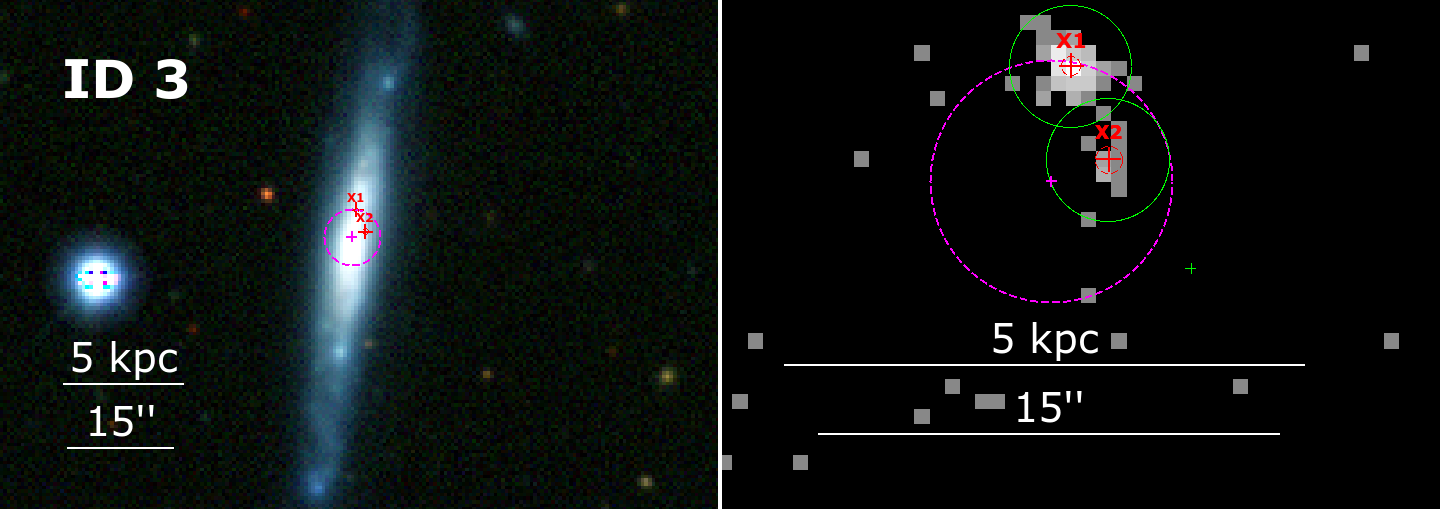}
\end{array}$
\end{center}
\caption{Left: Three-color images of our target galaxies retrieved from the Dark Energy Camera Legacy Survey (DECaLS; \citealt{2019AJ....157..168D}) with the $z$, $r$, and $g$ bands depicting red, green, and blue, respectively. Right: \textit{Chandra} X-ray images of the galaxies in the $0.5-7$~keV band. We show the positions and the corrected combined positional uncertainty (based on the \texttt{RADEC\_ERR\_CORR} parameter from the eFEDS main catalog) in magenta. We also overlay the positions of our detected X-ray sources as red crosses. The green circles represent the $\sim2\arcsec$ apertures around the sources. The red circle shows the 95\% positional uncertainties of the eROSITA-detected X-ray sources. Each \textit{Chandra}-detected source falls within the positional uncertainty reported in the eFEDS main catalog, except for ID~1, which does not have a detectable X-ray source in the \textit{Chandra} image.}
\label{fig:Images}
\end{figure*}

In this study, we present follow-up \textit{Chandra} X-ray observations of three nearby dwarf galaxies, each previously found to host one nuclear or off-nuclear X-ray source. These three targets come from a sample of six originally identified by \citet{2021ApJ...922L..40L}, who provided a
determination of the AGN fraction in dwarf galaxies
using X-ray observations from the eROSITA Final Equatorial Depth Survey (eFEDS; \citealt{2022A&A...661A...1B}). This work provided a meaningful example of what will be possible with the release of the eROSITA All-Sky Survey (eRASS) regarding the search for and understanding of AGNs in dwarf galaxies \citep{2024MNRAS.527.1962B,2024arXiv240601707S}.

\section{Sample Selection} \label{sec:sample}

Our target galaxies were selected from the study by \citet{2021ApJ...922L..40L}. They began with a parent sample of 63,582 local dwarf galaxies ($z \leqslant 0.15$, $M_{\star} \leqslant 3 \times 10^{9} \ M_{\odot}$) from the NASA-Sloan Atlas (v1\_0\_1) and found that 495 of these galaxies fall within the $\sim140$ deg$^{2}$ \citep{2022A&A...661A...1B} eFEDS field. They cross-matched the $\sim28,000$ X-ray sources in the eFEDS main catalog with the 495 NSA dwarf galaxies in the eFEDS footprint, resulting in a final sample of six dwarf galaxies with candidate massive BHs, each with one associated eFEDS X-ray source. We refer to \citet{2021ApJ...922L..40L} for further details on the sample.

All six of the galaxies exhibited moderately enhanced X-ray emission compared to their expected galaxy-wide contribution from XRBs based on the stellar mass and star formation rate of each galaxy, suggestive of accreting massive black holes (see Section 3.3 of \citealt{2021ApJ...922L..40L}). They found X-ray source luminosities of $L_{\rm 0.5-8keV} \sim 10^{39-40}$~erg~s$^{-1}$ across the galaxies in their sample.

Of the six dwarf galaxies, we have selected three (IDs 1, 2, 3) to obtain new {\it Chandra X-ray Observatory} observations {as part of the \textit{Chandra} GTO program} (PI: Kraft).  {\it Chandra} supplies us with a sub-arcsecond spatial resolution for targets located at the aimpoint, whereas the eROSITA resolution is $\sim16\arcsec$ \citep{2021A&A...647A...1P}. The much better resolution of \textit{Chandra} allows us to precisely constrain the locations and positional uncertainties of the X-ray sources previously detected in these galaxies, more accurately describe their X-ray emission, as well as fit spectra to determine further characteristics. We adopt star formation rates from \citet{2021ApJ...922L..40L},  based on the far-UV and mid-infrared luminosities via the relationships of \citet{2012ARA&A..50..531K} and \citet{2011ApJ...741..124H}. The properties of the three dwarf galaxies are summarized in Table \ref{tab:galprop}, and images with the positions of their associated X-ray source are in Figure \ref{fig:Images}.

\begin{deluxetable*}{lccccccccc}[t]
\tablewidth{0pt} 
\tablecaption{Galaxy Properties}
\tablehead{
\colhead{ID} & \colhead{NSAID} & \colhead{R.A.} & \colhead{Decl.} &
\colhead{$N_\text{H}$} & \colhead{$z$} & \colhead{$r_\text{50}$} & \colhead{Distance} & \colhead{log $M_{*}/M_{\sun}$} & \colhead{SFR}  \\
\colhead{} & \colhead{} & \colhead{(deg)} & \colhead{(deg)} & \colhead{($10^{20}$ cm$^{-2}$)} & 
\colhead{} & \colhead{(kpc)} & \colhead{(Mpc)} & \colhead{} & \colhead{($M_{\sun}$ yr$^{-1}$)} \\
\colhead{(1)} & \colhead{(2)} & \colhead{(3)} & \colhead{(4)} & \colhead{(5)} & \colhead{(6)} & \colhead{(7)} & \colhead{(8)} & \colhead{(9)} & \colhead{(10)} }
\startdata 
1 &  {82162} & {140.942967} & {2.753302} & {3.58} & {0.0177} &  {8.24}  & {73} & {9.08}  & {0.16} \\
2 & {623354} & {145.180051}  & {3.958864} & {3.64}  & {0.0051} & {0.91}   & {21}  & {8.40} & {0.04} \\
3 & {648474} & {141.572049} & {3.134800}  & {3.82} & {0.0149}  & {3.24} & {61}   & {9.44} & {0.27} \\
\enddata
\tablecomments{Table \ref{tab:galprop} information is taken from \citet{2021ApJ...922L..40L}. Column 1: Identification number used in \citet{2021ApJ...922L..40L}. Column 2: NSAIDs from v1\_0\_1 of the NSA. Columns 3 and 4: R.A. and Decl. of the galaxy. Column 5: Galactic neutral hydrogen density (retrieved via \url{https://cxc.harvard.edu/toolkit/colden.jsp}). Column 6: Redshift, specifically the \texttt{zdist} parameter from the NSA. Column 7: Petrosian 50\% light radius. Column 8: Galaxy distance. Column 9: Galaxy stellar mass. Column 10: Estimated SFRs from \citet{2021ApJ...922L..40L}. The values given in columns 6-9 are from the NSA and we assume \textit{h} = 0.73.}
\label{tab:galprop}
\end{deluxetable*}

\begin{deluxetable}{lcccc}[t]
\centering
\tablewidth{0pt} 
\tablecaption{Chandra Observations\label{tab:ChandraObs}}
\tablehead{
\colhead{ID} & \colhead{Date Observed} & \colhead{Obs ID} & \colhead{Exp. Time (ks)} & \colhead{$N_{\rm{bkg}}$} } 
\startdata 
1 & {2022 Oct 13}  & {26034} & {18.5}  & {1.3840} \\ 
  & {2022 Oct 16}  & {27479} & {11.4}  & {0.8680} \\
2 & {2022 Mar 07}  & {26035} & {12.9}  & {0.1456} \\ 
3 & {2023 Feb 27}  & {26036} & {22.8}  & {0.3561}  
\enddata
\tablecomments{\textit{N}$_{\rm{bkg}}$ is the expected number of hard band N ($>$S) background sources within our galaxy region (3r$_{50}$) using \citep{2003ApJ...588..696M}.}
\end{deluxetable}

\section{Observations and Data Reduction} \label{sec:obs}

The observations were taken with \textit{Chandra} between March 6, 2022, and February 27, 2023, with exposure times ranging between 11.4~ks and 22.8~ks. Each observation was taken with the galaxy centered on the S3 chip of the Advanced CCD Imaging Spectrometer (ACIS). A summary of the \textit{Chandra} observations is provided in Table \ref{tab:ChandraObs}. We carried out the data reduction using \texttt{CIAO} v.4.14 \citep{2006SPIE.6270E..1VF} software with the calibration dataset, CALDB v4.9.8.

We first reprocessed and calibrated the data using the \texttt{chandra\_repro} task, which resulted in new level 2 event files. Based on these event files, we generated images in the 0.5-2 keV (soft), 2-7 keV (hard), and 0.5-7 keV (broad) bands. To identify and remove any time intervals with high background periods, we used the \texttt{deflare} script and excluded time intervals where the background rate was $>3\sigma$ above the mean rate. For galaxy ID~1, two observations are available, which were merged using the \texttt{merge\_obs} task. 

To identify and create a list of X-ray point sources in the \textit{Chandra} images, we used the \texttt{wavdetect} tool in \texttt{CIAO} on the ACIS-S3 chip. We adopted wavelet scales of 1.0, 1.4, 2.0, 2.8, and 4.0 to search for point sources, setting the significance threshold to $10^{-6}$, which approximately results in one false source detection over the area of the S3 chip. While we detected multiple sources in each image, in the immediate vicinity of the galaxies, we detected zero, one, and two sources for galaxies ID~1, ID~2, and ID~3, respectively. The location, characteristics, and nature of these X-ray sources are further discussed in Section \ref{sec:analysis}.

\section{Analysis and Results} \label{sec:analysis}
\subsection{X-ray Sources}
\label{subsec:sources}
 
We begin by determining if any X-ray sources are associated with the target galaxies by searching the regions within $3r_{\rm 50}$, where $r_{50}$ is the half-light radius provided by the NSA. For galaxy ID~1, the detected X-ray sources lie well beyond the optical extent of the galaxy in both the individual and merged observations. For galaxies ID~2 and ID~3, we detect X-ray sources within the optical extent of the galaxies with one source in ID~2 and two sources in ID~3. Based on the locations of the X-ray sources, we find that their positions are consistent with the X-ray source positions and positional uncertainties of the sources detected by eROSITA \citep{2021ApJ...922L..40L}.

Figure~\ref{fig:Images} shows the DECaLS optical images and the {\it Chandra} X-ray images of the three galaxies. This reveals that the X-ray source detected in ID~2 is relatively central, with an offset of $\approx1\arcsec$ from its poorly defined center. For galaxy ID~3, one of the sources, X2, is roughly associated with the nucleus of the dwarf galaxy, with an offset of $\approx2.8\arcsec$, while the offset between the northern source, X1, and the galaxy centroid is $\approx5\arcsec$.

To derive the number of X-ray counts associated with the sources, we applied circular apertures with a $2\arcsec$ radius, which encompass $90\%$ of the enclosed energy fraction at 4.5~keV. The background counts were derived from annuli with inner and outer radii of $2\arcsec-24\arcsec$. We derived the number of net counts for each source region and applied a $90\%$ aperture correction.

We used the \texttt{CIAO} \texttt{srcflux} tool to derive the X-ray fluxes of the sources in the $0.5-2$~keV and $2-10$~keV  bands. We assumed a power-law spectral model with a photon index of $\Gamma=1.8$, which is common for low-luminosity AGN \citep{2008ARA&A..46..475H, 2009ApJ...699..626H}, and ultraluminous X-ray sources in this luminosity range \citep{2008ApJ...684..282S}. We use the Galactic column density maps from \citet{1990ARA&A..28..215D}. We note that potential absorption intrinsic to each source is not accounted for and therefore the observed fluxes should be considered as lower limits.

To derive the luminosities of the sources, we use the distances in Table \ref{tab:galprop}. The obtained $2-10$ keV band luminosities are in the range of $\log (L_{\rm 2-10 keV}$/erg s$^{-1}) = 39.20-40.43$. We report the number of counts, the unabsorbed fluxes and luminosities in Table \ref{tab:xraysrc}. We note that ID 3 - X1 has a luminosity $\sim 5\times$ higher than the eFEDS source detected in ID 3, while the other detected X-ray sources are consistent with the eFEDS luminosities. 

In the absence of a detection in the galaxy ID 1, we place an upper limit on the luminosity of any potential source {at the location of the eFEDS detection. The upper limit from Chandra is $L_{\rm 0.5-2 keV} < 2.4 \times 10^{-15}$ erg s$^{-1}$ cm$^{-2}$ and the eFEDS source in ID 1 has $L_{\rm 0.5-2 keV} = 2.3 \times 10^{-15}$ erg s$^{-1}$ cm$^{-2}$ \citep{2021ApJ...922L..40L}. Given the consistency of these values, the source could be relatively soft and not detected by \textit{Chandra} due to decreased sensitivity in the soft band, or the source could be variable. While spectral analysis could help differentiate between these scenarios, the eROSITA-detected source is too faint to carry out such an analysis.
}

\begin{deluxetable*}{ccccccccc}
\tabletypesize{\footnotesize}
\tablecaption{X-ray Sources \label{tab:xraysrc}}
\tablewidth{0pt}
\tablehead{
\colhead{ID} & \colhead{R.A.} & \colhead{Decl.} & \multicolumn{2}{c}{Net Counts} & \multicolumn{2}{c}{Flux ($10^{-15}$ erg s$^{-1}$ cm$^{-2}$)} & \multicolumn{2}{c}{Luminosity (log(erg s$^{-1}$))} \\
\cmidrule(l){4-5} \cmidrule(l){6-7} \cmidrule(l){8-9}
\colhead{ } & \colhead{(deg)} & \colhead{(deg)} & \colhead{0.5-2 keV} & \colhead{2-7 keV} & \colhead{0.5-2 keV} & \colhead{2-10 keV} & \colhead{0.5-2 keV} & \colhead{2-10 keV} \\
\colhead{(1)} & \colhead{(2)} & \colhead{(3)} & \colhead{(4)} & \colhead{(5)} & \colhead{(6)} & \colhead{(7)} & \colhead{(8)} & \colhead{(9)}}
\startdata
1 
& \nodata & \nodata & \nodata  & \nodata & $< 2.40$   & $< 5.23$   & $< 39.2$    & $<39.5$  \\
2 	 & {145.179013} & {3.959214}  & {28.5$\pm 9.7$} & 15.3$^{+8.7}_{-6.1}$  & 27.43    & 29.72    & 39.16    & 39.20  \\
3-X1 & {141.571720} & {3.136256} & {57.3$\pm 13.5$}  & {58.1$\pm 13.8$} & {32.58}    & {59.89}    & {40.16}    & {40.43}  \\
3-X2 & {141.571383} & {3.135414} & 10.3$^{+7.3}_{-4.8}$    & \nodata   & 5.88 & \nodata & 39.42 & \nodata
\enddata 
\tablecomments{Column 1: galaxy ID. 
Columns 2 and 3:  R.A. and Decl. of the X-ray sources.  
Columns 4 and 5: Net counts after applying a 90\% aperture correction.  Error bars represent 90\% confidence intervals.
Columns 6 and 7: Fluxes corrected for Galactic absorption.
Columns 8 and 9: Luminosities corrected for Galactic absorption; calculated assuming a photon index of $\Gamma = 1.8$.
}
\end{deluxetable*}

\subsection{Spectral Properties}

\begin{figure}[!t]
    \centering
    \plotone{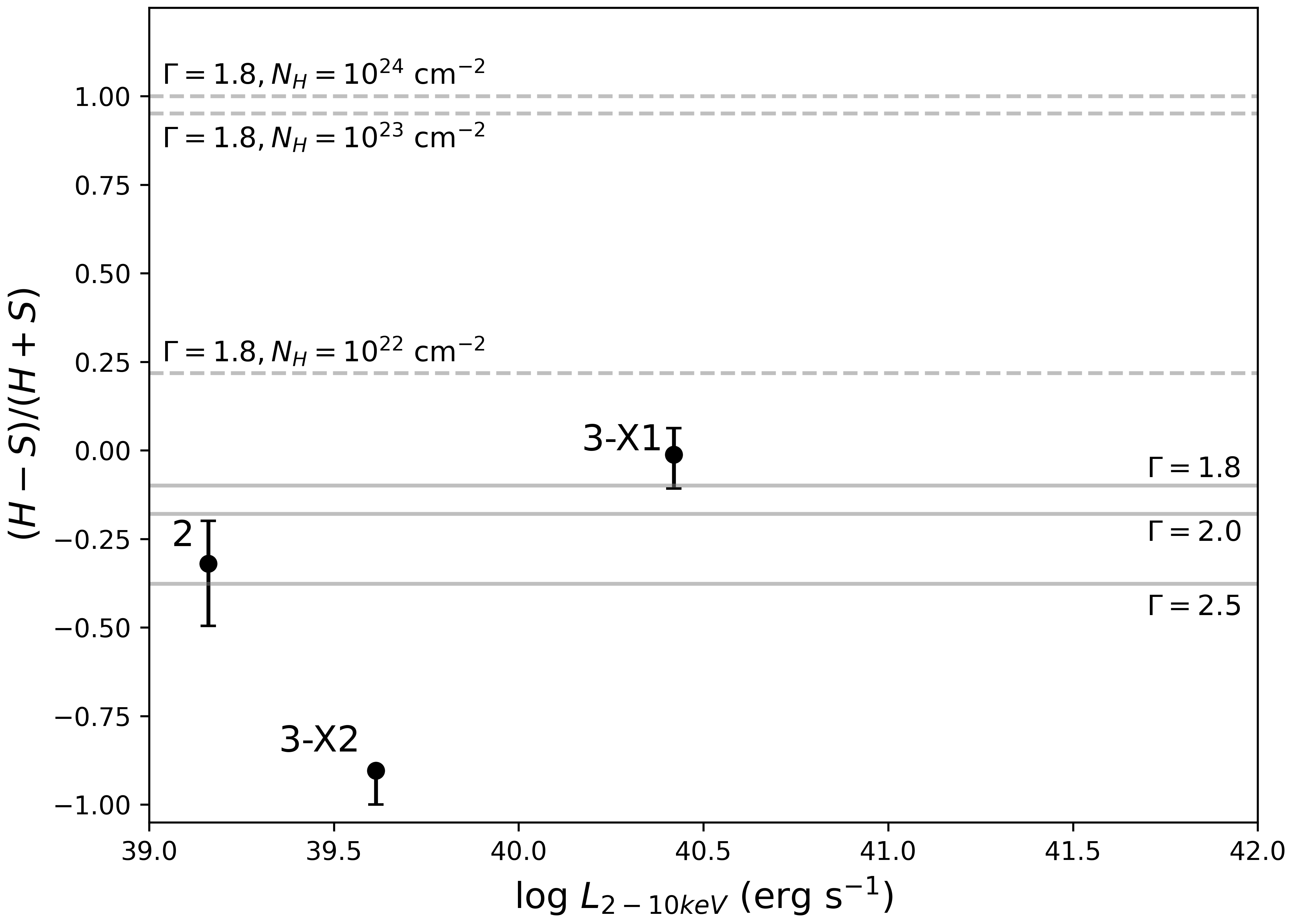}
    \caption{Hardness ratio vs. $2–10$~keV band X-ray luminosity for our two galaxies with a detected X-ray source in at least one of the two bands. The hardness ratio was calculated using BEHR (see Section \ref{subsec:sources}). The error bars are the 68\% confidence intervals.  Hardness ratios for unabsorbed power law models with $\Gamma=$ 1.8, 2.0, and 2.5 are depicted as solid gray lines, while the absorbed  power laws with $\Gamma=1.8$ are shown with gray dashed lines.}
    \label{fig:HR}
\end{figure}

We derived the X-ray hardness ratios using the Bayesian Estimation of Hardness Ratios (BEHR) code \citep{2006ApJ...652..610P}. BEHR is a robust tool for estimating hardness ratios for sources within the Poisson regime of low counts, which also provides uncertainties if a source is detected in only one of the two bands. We define the hardness ratio as $HR = (H-S)/(H+S)$, where $H$ and $S$ are the numbers of detected counts in the hard ($2-7$~keV) and soft ($0.5-2$~keV) bands, respectively. We note that source X2 in galaxy ID3 is not detected in the hard band; hence, we apply a correction factor of $\sim1.6$ to the soft band flux using the spectral models introduced below. The resulting hardness ratios are in the range of $HR = -0.9 - 0$ and are shown in Figure \ref{fig:HR}.

To place these hardness ratios in context, we derived the expected values for a range of spectral models. To this end, we used the Portable, Interactive Multi-Mission Simulator (PIMMS)\footnote{\url{https://heasarc.gsfc.nasa.gov/cgi-bin/Tools/w3pimms/w3pimms.pl}} tool and derived hardness ratios for unabsorbed power-law models with $\Gamma = 1.8$, $\Gamma = 2.0$, and $\Gamma = 2.5$. Additionally, we derived the expected hardness ratios assuming a power-law model with a slope of $\Gamma = 1.8$ and intrinsic absorption of $N_H = 10^{22}$ cm$^{-2}$, $N_H = 10^{23}$ cm$^{-2}$, and $N_H = 10^{24}$ cm$^{-2}$. {These are shown as solid and dashed lines respectively in Figure \ref{fig:HR}.}

We find that the observed hardness ratios are broadly consistent with unabsorbed power-law models with $\Gamma = 1.8-2.5$, although source X2 in galaxy ID~3 exhibits a significantly softer spectrum, which is also indicated by its non-detection in the hard band.

\begin{figure*}[t!]    
    \centering
        \plottwo{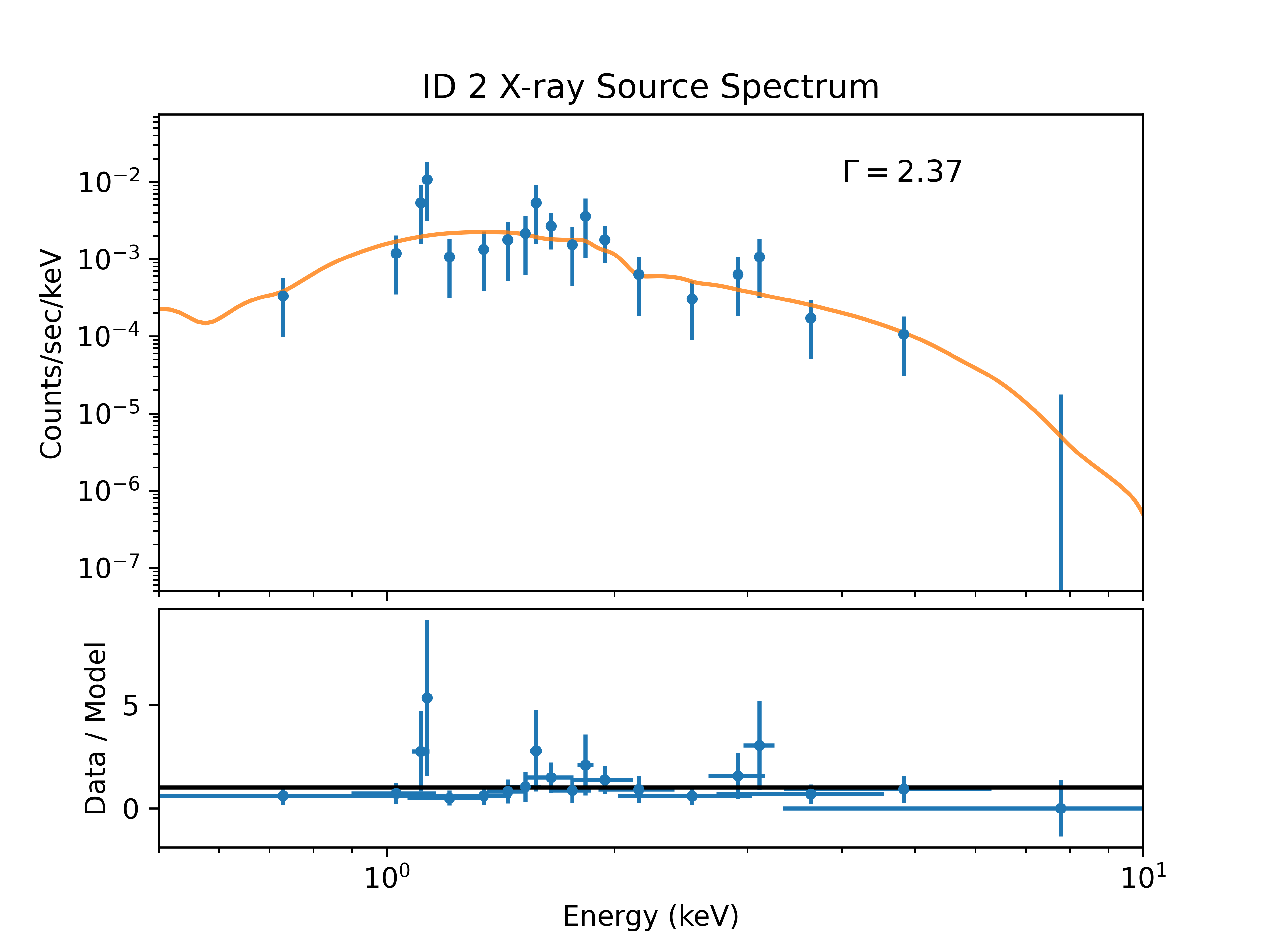}{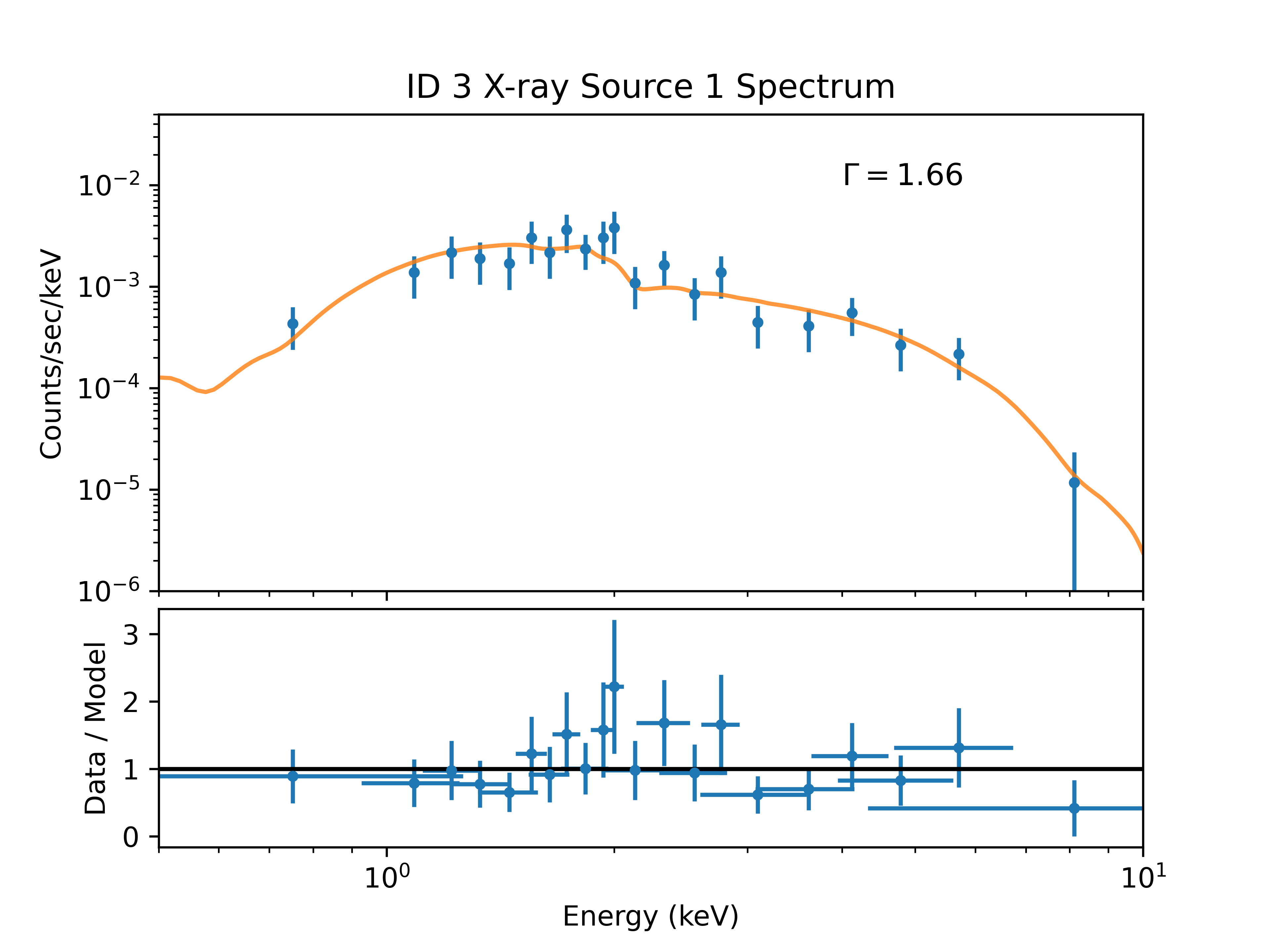}
\caption{X-ray energy spectra for the X-ray source detected in galaxy ID~2 (left panel) and for source X1 in ID~3 (right panel). The spectra were fit with absorbed power-law models, resulting in best-fit values of $\Gamma = 2.37^{+0.33}_{-0.32}$ and $\Gamma = 1.66^{+0.25}_{-0.19}$ for the sources in ID~2 and ID~3, respectively.} 
\label{fig:spectra}
\end{figure*}

We also modeled and fitted the spectra of two of our X-ray sources, the source in ID~2 and source X1 in ID~3, with \texttt{CIAO}'s built-in modeling and fitting package, \texttt{SHERPA} v.4.14 \citep{2021zndo...5554957B}. The extracted source and background apertures were the same as described earlier (see Section \ref{sec:obs}). We used the \texttt{CIAO} \texttt{specextract} tool to extract the spectra and the response files. Since our sources contain a relatively low number of photon counts, we used the C-statistic to calculate the best-fitting parameters. We modeled the background along with our source spectra.

Given the possibility of both Galactic absorption and absorption intrinsic ($N_{H,int}$) to the target galaxies, we fitted the spectra using two models: (i) an absorbed power-law model without intrinsic absorption (\texttt{xsphabsxspowerlaw}) and (ii) an absorbed power-law with intrinsic absorption (\texttt{xsphabsxsphabs*xspowerlaw}). For both models, we used the Galactic column densities reported by \citet{1990ARA&A..28..215D}, which we kept fixed. We allowed all other parameters to vary and estimated all uncertainties to $68\%$ confidence.

Performing the fit for each case on our sources, we found that the spectrum of ID~2 has a best-fit power-law index of $\Gamma = 2.37^{+0.33}_{-0.32}$. There was no significant difference between the two models for the spectrum of ID~2, as it yielded a negligible value for the best-fit intrinsic absorption. For ID~3, the X1 X-ray source was the only detection with sufficient counts for spectral fitting. We found a similar situation, with both models yielding a best-fit power-law index of $\Gamma = 1.66^{+0.25}_{-0.19}$, with the second model arriving at a negligible intrinsic absorption. We show the spectrum for each galaxy and their best-fit power-law models in Figure~\ref{fig:spectra}. We calculated unabsorbed $2-10$ keV X-ray source fluxes and luminosities based on the modeled spectra and summarized the results in Table~\ref{tab:erossrc}.

We note that the best-fit power-law indices for the source in ID~2 and source X1 in ID~3 are consistent with the hardness ratios calculated above, with $\Gamma \approx 2.4$ and $\Gamma \approx 1.7$, respectively (see Figure~\ref{fig:HR}). The X-ray source luminosities found, whether via \texttt{srcflux} or fitting the spectra, return values that are also consistent with values found in other studies concerning AGNs in dwarf galaxies \citep[e.g.,][]{2014ApJ...787L..30R, 2017ApJ...836...20B, 2018MNRAS.478.2576M}.

\begin{deluxetable*}{ccccc}[t!]    
\tablewidth{0pt} \label{tab:latcomp}
\tablenum{4}
\tablecaption{X-ray Source Properties From Fitting\label{tab:erossrc}}
\tablehead{
\colhead{ID}  & \colhead{\textit{C}-stat/d.o.f.} & \colhead{$\Gamma$} & \colhead{\textit{F}$_{\text{2-10 keV}}$ } & \colhead{log \textit{L}$_{\text{2-10 keV}}$} \\
\colhead{} & \colhead{} & \colhead{} & \colhead{(10$^{-15}$ erg s$^{-1}$ cm$^{-2}$)} & \colhead{(erg s$^{-1}$)} \\
 {(1)} & {(2)} & {(3)} & {(4)} & {(5)}}
\startdata 
{2} & {24.6/30} & {2.37$\substack{+0.33 \\ -0.32}$} & {21.97$\substack{+15.16 \\ -8.98}$}  & {39.1$\pm$ 0.2} \\
{3-X1} & {50/48} & {1.66$\substack{+0.25 \\ -0.19}$}  & {60.27$\substack{+34.80 \\ -22.52}$} & {40.4$\pm$0.2} 
\enddata
\tablecomments{Column 1: Identification number associated with the X-ray source. Column 2: Goodness-of-fit/degrees of freedom. Column 2: Best-fit photon index.  Column 3: Fluxes in the $2-10$~keV band found from the modeled spectra corrected for absorption. Column 4: Luminosity in $2-10$~keV band. {The Galactic $N_{\rm{H}}$ values given in Table \ref{tab:galprop} provided the best fit for both sources.}}
\end{deluxetable*}

\section{Discussion}

\subsection{Possible Origins of the X-ray Emission}

Below we consider a number of possible origins for the X-ray sources we detect towards our target dwarf galaxies. Based on the eFEDS detections, \citet{2021ApJ...922L..40L} found luminosities above the galaxy-wide expected contribution from XRBs. Here we consider contamination from the cosmic X-ray background (CXB) and ultraluminous X-ray sources (ULXs).

\subsubsection{Contamination from CXB Sources}

We first estimate the number of hard/soft X-ray sources from the CXB that we expect may reside within 3\textit{r}$_{50}$ of the target galaxies.
Using the $\log N - \log S$ function established by  \citet{2003ApJ...588..696M}, we derive the number of expected CXB sources per unit area given a limiting flux. Based on the source detection threshold achieved by the \textit{Chandra} observations for our target galaxies, the faintest detectable sources are in the range of $S_{\rm min,soft} = (1.2-2) \times 10^{-15} \ \rm{erg \ s^{-1} \ cm^{-2}}$ and $S_{\rm min,hard} = (2.3-4) \times 10^{-15} \ \rm{erg \ s^{-1} \ cm^{-2}}$  for the $0.5-2$~keV and $2-10$~keV bands, respectively. Based on these fluxes, the $\log N - \log S$ function, and the area of the 3\textit{r}$_{50}$ for the galaxies, we expect to detect $0.085-0.525$ and $0.147-1.38$ CXB sources in the soft and hard bands, respectively. Therefore, it is unlikely that all detected sources are resolved CXB sources. {The location of source X1 in ID 1 lies far to the northwest of the target galaxy (see Figure \ref{fig:Images}) yet lies within 3\textit{r}$_{50}$ ($\sim 24$ kpc; Table \ref{tab:galprop}). It is therefore a prime candidate for a background source.}

\subsubsection{Ultraluminous X-ray Sources}\label{sec:ulx}

Next, we take into consideration the likelihood of ultraluminous X-ray sources (ULXs) as the origin of the X-ray sources detected in our sample. ULXs are described as being off-nuclear sources with X-ray luminosities above 10$^{39}$ erg s$^{-1}$ (see \citealt{2017ARA&A..55..303K} for a review). This population is consistent with the extension of the bright end tail of the luminosity function of high-mass X-ray binaries.  Our source positions and their luminosities fall squarely into that definition. We refer to the analysis done by \citet{2021ApJ...922L..40L}, who consider various relations to estimate the expected number of ULXs in their sample of six dwarf galaxies. Ultimately, they conclude that $\sim 1$ of the X-ray sources could be a ULX. Taking into account our smaller sample size and hence lower combined star-formation rate, we expect that our sample contains $\sim0.5$ ULXs. 

Given that we detect two X-ray sources in ID 3 with our {\it Chandra} observations, it is plausible that one of the sources is a ULX and one is an AGN. {While we cannot definitively say which source is which, our favored interpretation is that X1 is a ULX while X2 is an AGN. First, in contrast to X2, X1 is more offset with respect to the previous eFEDS detection and has a position just barely consistent with the optical center of the galaxy. It is also highly variable with a Chandra luminosity $\sim 5\times$ higher than the eFEDS luminosity (\S\ref{subsec:sources}), whereas source X2 has a Chandra luminosity consistent with the eFEDS luminosity. While variability is not inconsistent with AGN activity, this behavior is quite common for stellar mass BHs. The photon index ($\Gamma \sim 1.7$) derived from the spectrum of X1 is also entirely consistent with known ULXs \citep{2004ApJS..154..519S}. Moreover, the softer value of the hardness ratio for X2 is consistent with previous X-ray studies of AGNs in dwarf galaxies \citep{2016ApJ...830L..35R,2017ApJ...836...20B}.}

\subsection{Massive Black Holes}

If the X-ray luminosities we have measured are due to accreting massive BHs, we can estimate their Eddington ratios.
The Eddington ratio of an accreting BH is found via 

\begin{equation}
f_{\rm Edd} = (\kappa L_{\rm X})/(L_{\rm Edd})
\end{equation}

\noindent
where $\kappa$ is the hard ($2-10$~keV) bolometric correction, $L_{\rm X}$ is the hard X-ray luminosity of the BH, and $L_{\rm Edd}$ is the Eddington luminosity of the BH, which is given by

\begin{equation}
L_{\rm Edd} \approx 1.26 \times 10^{38} \frac{M_{\rm BH}}{M_\odot} \text{ erg s$^{-1}$.}\end{equation}

\noindent
The BH masses are estimated using Equation (4) from \citet{2015ApJ...813...82R}, which gives log($M_{\rm BH})$ as a function of total host galaxy stellar mass, log($M_{*}$). The relation for local AGNs is given by:

\begin{equation}
\text{log}(M_{\text{BH}}/M_{\odot}) = \alpha + \beta \text{log}(M_{*}/10^{11}M_{\odot}),  
\end{equation}
where $\alpha = 7.45 \pm 0.08$ and $\beta = 1.05 \pm 0.11$.
The resultant BH mass estimates are log($M_{\rm BH}/M_\odot$) $\approx$ 4.7 and 5.8 for the AGNs in ID~2 and ID~3, respectively. These BH masses carry uncertainties of $\sim 0.55$ dex \citep{2015ApJ...813...82R}.

To apply the hard band bolometric correction, we invoke \citet{2020A&A...636A..73D} that returns a mean $L_{\rm bol}/L_{\rm 2-10 keV} \approx 15.3$. 
Combining these values, we find Eddington fractions of 
$\sim$0.3\% and $\sim$0.08\% for the candidate AGNs in ID~2 and ID 3 (X2), respectively.

\subsection{Comparison with \citet{2021ApJ...922L..40L}}

To compare the X-ray luminosities from eROSITA presented by \citet{2021ApJ...922L..40L} with our values from {\it Chandra}, we must adopt the same parameters in the analysis. To be consistent with \citet{2021ApJ...922L..40L}, we re-estimate source fluxes using a photon index of $\Gamma$ = 2.0 and Galactic absorption of 3.0 $\times$ 10$^{20}$ cm$^{-2}$. We also run through our analysis in the $0.5-8$~keV band. To compare ID~3-X2, we used the above parameters and analyzed them to obtain soft band fluxes, which we convert to $0.5-8$~keV using a correction factor of $\sim2.1$ found through PIMMS.

The luminosities of the X-ray source in ID 2 and the source X2 in ID~3 are consistent with the sources \citet{2021ApJ...922L..40L} originally detected in the eROSITA data. However, we also detect an additional source in ID 3 (X1) that is a bit offset from the nucleus with a luminosity of log (\textit{L}$_{\text{2-10 keV}}$/erg s$^{-1}$) $\approx 40.4$. Given the $16\arcsec$ resolution of eROSITA \citep{2021A&A...647A...1P}, X1 and X2 in ID 3 would not have been resolved in the eFEDS catalog since we find a spatial offset between X1 and X2 of $\sim$3$''$ ( $\sim$1 kpc projected). It may be that X2 was the source originally detected by \citet{2021ApJ...922L..40L} and we have detected a separate, new source that could be a ULX (see \S\ref{sec:ulx}).

\section{Summary and Conclusions}

We analyze new {\it Chandra X-ray Observatory} observations of three  AGN candidates in dwarf galaxies identified by Latimer et al.\ (\citeyear{2021ApJ...922L..40L}) from the eFEDS catalog. Our main findings are summarized below:

\begin{itemize}

\item X-ray emission is detected in two of our three target galaxies (one source in ID~2 and two sources in ID~3) with the X-ray sources being located within the positional uncertainty reported by eFEDS. We do not detect a luminous X-ray source in ID~1 within the positional uncertainty reported by eFEDS nor the optical extent of the galaxy itself. 

\item The X-ray source in ID 2 has a luminosity consistent with the eFEDS source ($L_{\text{2-10 keV}} \sim 1 \times 10^{39}$ erg s$^{-1}$) and has a position that is roughly central within the target galaxy, although there does not seem to be a well-defined center. Given the luminosity is also well above that expected from XRBs based on the host galaxy stellar mass and SFR \citep{2021ApJ...922L..40L}, the X-ray source may indeed be an accreting massive BH. The X-ray spectrum is well-described by an absorbed power-law model with $\Gamma = 2.4$ and negligible intrinsic absorption. The soft spectrum is consistent with other X-ray studies of AGNs in dwarf galaxies \citep[e.g.,][]{2016ApJ...830L..35R, 2017ApJ...836...20B}. 

\item Our {\it Chandra} observations resolved the previous eFEDS source in ID 3 into two individual sources. X2 has a luminosity consistent with that in eFEDS ($L_{\rm 2-10 keV} \sim 4 \times 10^{39}$ erg s$^{-1}$), which is higher than that expected from XRBs \citep{2021ApJ...922L..40L}. X2 also has a position approximately consistent with the optical nucleus of the galaxy, strongly suggesting this X-ray source is due to an active massive BH.  X1 has a luminosity $\sim 5\times$ higher than the eFEDS source detected in ID 3 ($L_{\rm 2-10 keV} \sim 2 \times 10^{40}$ erg s$^{-1}$) and is offset from the nucleus. The X-ray spectrum of X1 is best fit by a power law model with a photon index of $\Gamma=1.7$ and negligible intrinsic absorption.  Given its position and high variability, this source is potentially a ULX.

\item For ID 2 and ID3-X2, BH masses are estimated using the scaling between BH mass and host galaxy stellar mass for local AGNs from \citet{2015ApJ...813...82R}. The BH mass estimates are log($M_{\rm BH}/M_\odot$) $\approx$ 4.7 and 5.8 for ID 2 and ID3-X2, respectively and carry uncertainties of $\sim 0.55$ dex. The corresponding Eddington ratios are $\sim$0.3\% for ID 2 and $\sim$ 0.08\% for ID3-X2.

\end{itemize}

Through this work, we have strengthened the case for the presence of massive BHs in two of the dwarf galaxies (IDs 2 and 3) identified by \citet{2021ApJ...922L..40L}, while weakening the case for a highly offset AGN in ID~1. Additional follow-up observations at radio wavelengths would help confirm the sources in ID~2 and ID~3 are indeed AGNs since massive BHs are much more luminous than XRBs in the radio.  

\section*{Acknowledgments}
{We thank the reviewer for their constructive comments.} A.E.R. gratefully acknowledges support for this work provided by the NSF through CAREER award 2235277 and NASA through EPSCoR grant No. 80NSSC20M0231. \'A.B. and R.P.K. acknowledge support from the Smithsonian Institution and the Chandra Project through NASA contract NAS8-03060. This paper employs a list of Chandra datasets, obtained by the Chandra X-ray Observatory, contained in~\dataset[DOI: 10.25574/cdc.282]{https://doi.org/10.25574/cdc.282}.


\begin{thebibliography}{}
\expandafter\ifx\csname natexlab\endcsname\relax\def\natexlab#1{#1}\fi
\providecommand{\url}[1]{\href{#1}{#1}}
\providecommand{\dodoi}[1]{doi:~\href{http://doi.org/#1}{\nolinkurl{#1}}}
\providecommand{\doeprint}[1]{\href{http://ascl.net/#1}{\nolinkurl{http://ascl.net/#1}}}
\providecommand{\doarXiv}[1]{\href{https://arxiv.org/abs/#1}{\nolinkurl{https://arxiv.org/abs/#1}}}

\bibitem[{{Baldassare} {et~al.}(2017){Baldassare}, {Reines}, {Gallo}, \&
  {Greene}}]{2017ApJ...836...20B}
{Baldassare}, V.~F., {Reines}, A.~E., {Gallo}, E., \& {Greene}, J.~E. 2017,
  \apj, 836, 20, \dodoi{10.3847/1538-4357/836/1/20}

\bibitem[{{Birchall} {et~al.}(2020){Birchall}, {Watson}, \&
  {Aird}}]{2020MNRAS.492.2268B}
{Birchall}, K.~L., {Watson}, M.~G., \& {Aird}, J. 2020, \mnras, 492, 2268,
  \dodoi{10.1093/mnras/staa040}

\bibitem[{{Brunner} {et~al.}(2022){Brunner}, {Liu}, {Lamer}, {Georgakakis},
  {Merloni}, {Brusa}, {Bulbul}, {Dennerl}, {Friedrich}, {Liu}, {Maitra},
  {Nandra}, {Ramos-Ceja}, {Sanders}, {Stewart}, {Boller}, {Buchner}, {Clerc},
  {Comparat}, {Dwelly}, {Eckert}, {Finoguenov}, {Freyberg}, {Ghirardini},
  {Gueguen}, {Haberl}, {Kreykenbohm}, {Krumpe}, {Osterhage}, {Pacaud},
  {Predehl}, {Reiprich}, {Robrade}, {Salvato}, {Santangelo}, {Schrabback},
  {Schwope}, \& {Wilms}}]{2022A&A...661A...1B}
{Brunner}, H., {Liu}, T., {Lamer}, G., {et~al.} 2022, \aap, 661, A1,
  \dodoi{10.1051/0004-6361/202141266}

\bibitem[{{Burke} {et~al.}(2021){Burke}, {Laurino}, {Wmclaugh}, {Dtnguyen2},
  {G{\"u}nther}, {Marie-Terrell}, {Siemiginowska}, {Budynkiewicz}, {Aldcroft},
  {Deil}, {Sip{\H{o}}cz}, {Buchner}, {Laginja}, {Leinweber}, {Nplee}, \&
  {Todd}}]{2021zndo...5554957B}
{Burke}, D., {Laurino}, O., {Wmclaugh}, {et~al.} 2021, {sherpa/sherpa: Sherpa
  4.14.0}, 4.14.0, Zenodo,  Zenodo, \dodoi{10.5281/zenodo.5554957}

\bibitem[{{Bykov} {et~al.}(2024){Bykov}, {Gilfanov}, \&
  {Sunyaev}}]{2024MNRAS.527.1962B}
{Bykov}, S.~D., {Gilfanov}, M.~R., \& {Sunyaev}, R.~A. 2024, \mnras, 527, 1962,
  \dodoi{10.1093/mnras/stad3355}

\bibitem[{{Dey} {et~al.}(2019){Dey}, {Schlegel}, {Lang}, {Blum}, {Burleigh},
  {Fan}, {Findlay}, {Finkbeiner}, {Herrera}, {Juneau}, {Landriau}, {Levi},
  {McGreer}, {Meisner}, {Myers}, {Moustakas}, {Nugent}, {Patej}, {Schlafly},
  {Walker}, {Valdes}, {Weaver}, {Y{\`e}che}, {Zou}, {Zhou}, {Abareshi},
  {Abbott}, {Abolfathi}, {Aguilera}, {Alam}, {Allen}, {Alvarez}, {Annis},
  {Ansarinejad}, {Aubert}, {Beechert}, {Bell}, {BenZvi}, {Beutler}, {Bielby},
  {Bolton}, {Brice{\~n}o}, {Buckley-Geer}, {Butler}, {Calamida}, {Carlberg},
  {Carter}, {Casas}, {Castander}, {Choi}, {Comparat}, {Cukanovaite}, {Delubac},
  {DeVries}, {Dey}, {Dhungana}, {Dickinson}, {Ding}, {Donaldson}, {Duan},
  {Duckworth}, {Eftekharzadeh}, {Eisenstein}, {Etourneau}, {Fagrelius},
  {Farihi}, {Fitzpatrick}, {Font-Ribera}, {Fulmer}, {G{\"a}nsicke},
  {Gaztanaga}, {George}, {Gerdes}, {Gontcho}, {Gorgoni}, {Green}, {Guy},
  {Harmer}, {Hernandez}, {Honscheid}, {Huang}, {James}, {Jannuzi}, {Jiang},
  {Joyce}, {Karcher}, {Karkar}, {Kehoe}, {Kneib}, {Kueter-Young}, {Lan},
  {Lauer}, {Le Guillou}, {Le Van Suu}, {Lee}, {Lesser}, {Perreault Levasseur},
  {Li}, {Mann}, {Marshall}, {Mart{\'\i}nez-V{\'a}zquez}, {Martini}, {du Mas des
  Bourboux}, {McManus}, {Meier}, {M{\'e}nard}, {Metcalfe},
  {Mu{\~n}oz-Guti{\'e}rrez}, {Najita}, {Napier}, {Narayan}, {Newman}, {Nie},
  {Nord}, {Norman}, {Olsen}, {Paat}, {Palanque-Delabrouille}, {Peng},
  {Poppett}, {Poremba}, {Prakash}, {Rabinowitz}, {Raichoor}, {Rezaie},
  {Robertson}, {Roe}, {Ross}, {Ross}, {Rudnick}, {Safonova}, {Saha},
  {S{\'a}nchez}, {Savary}, {Schweiker}, {Scott}, {Seo}, {Shan}, {Silva},
  {Slepian}, {Soto}, {Sprayberry}, {Staten}, {Stillman}, {Stupak}, {Summers},
  {Sien Tie}, {Tirado}, {Vargas-Maga{\~n}a}, {Vivas}, {Wechsler}, {Williams},
  {Yang}, {Yang}, {Yapici}, {Zaritsky}, {Zenteno}, {Zhang}, {Zhang}, {Zhou}, \&
  {Zhou}}]{2019AJ....157..168D}
{Dey}, A., {Schlegel}, D.~J., {Lang}, D., {et~al.} 2019, \aj, 157, 168,
  \dodoi{10.3847/1538-3881/ab089d}

\bibitem[{{Dickey} \& {Lockman}(1990)}]{1990ARA&A..28..215D}
{Dickey}, J.~M., \& {Lockman}, F.~J. 1990, \araa, 28, 215,
  \dodoi{10.1146/annurev.aa.28.090190.001243}

\bibitem[{{Duras} {et~al.}(2020){Duras}, {Bongiorno}, {Ricci}, {Piconcelli},
  {Shankar}, {Lusso}, {Bianchi}, {Fiore}, {Maiolino}, {Marconi}, {Onori},
  {Sani}, {Schneider}, {Vignali}, \& {La Franca}}]{2020A&A...636A..73D}
{Duras}, F., {Bongiorno}, A., {Ricci}, F., {et~al.} 2020, \aap, 636, A73,
  \dodoi{10.1051/0004-6361/201936817}

\bibitem[{{Fruscione} {et~al.}(2006){Fruscione}, {McDowell}, {Allen},
  {Brickhouse}, {Burke}, {Davis}, {Durham}, {Elvis}, {Galle}, {Harris},
  {Huenemoerder}, {Houck}, {Ishibashi}, {Karovska}, {Nicastro}, {Noble},
  {Nowak}, {Primini}, {Siemiginowska}, {Smith}, \&
  {Wise}}]{2006SPIE.6270E..1VF}
{Fruscione}, A., {McDowell}, J.~C., {Allen}, G.~E., {et~al.} 2006, in Society
  of Photo-Optical Instrumentation Engineers (SPIE) Conference Series, Vol.
  6270, Observatory Operations: Strategies, Processes, and Systems, ed. D.~R.
  {Silva} \& R.~E. {Doxsey}, 62701V, \dodoi{10.1117/12.671760}

\bibitem[{{Gallo} {et~al.}(2008){Gallo}, {Treu}, {Jacob}, {Woo}, {Marshall}, \&
  {Antonucci}}]{2008ApJ...680..154G}
{Gallo}, E., {Treu}, T., {Jacob}, J., {et~al.} 2008, \apj, 680, 154,
  \dodoi{10.1086/588012}

\bibitem[{{Greene} {et~al.}(2020){Greene}, {Strader}, \&
  {Ho}}]{2020ARA&A..58..257G}
{Greene}, J.~E., {Strader}, J., \& {Ho}, L.~C. 2020, \araa, 58, 257,
  \dodoi{10.1146/annurev-astro-032620-021835}

\bibitem[{{Hainline} {et~al.}(2016){Hainline}, {Reines}, {Greene}, \&
  {Stern}}]{2016ApJ...832..119H}
{Hainline}, K.~N., {Reines}, A.~E., {Greene}, J.~E., \& {Stern}, D. 2016, \apj,
  832, 119, \dodoi{10.3847/0004-637X/832/2/119}

\bibitem[{{Hao} {et~al.}(2011){Hao}, {Kennicutt}, {Johnson}, {Calzetti},
  {Dale}, \& {Moustakas}}]{2011ApJ...741..124H}
{Hao}, C.-N., {Kennicutt}, R.~C., {Johnson}, B.~D., {et~al.} 2011, \apj, 741,
  124, \dodoi{10.1088/0004-637X/741/2/124}

\bibitem[{{Hickox} {et~al.}(2009){Hickox}, {Jones}, {Forman}, {Murray},
  {Kochanek}, {Eisenstein}, {Jannuzi}, {Dey}, {Brown}, {Stern}, {Eisenhardt},
  {Gorjian}, {Brodwin}, {Narayan}, {Cool}, {Kenter}, {Caldwell}, \&
  {Anderson}}]{2009ApJ...696..891H}
{Hickox}, R.~C., {Jones}, C., {Forman}, W.~R., {et~al.} 2009, \apj, 696, 891,
  \dodoi{10.1088/0004-637X/696/1/891}

\bibitem[{{Ho}(2008)}]{2008ARA&A..46..475H}
{Ho}, L.~C. 2008, \araa, 46, 475,
  \dodoi{10.1146/annurev.astro.45.051806.110546}

\bibitem[{{Ho}(2009)}]{2009ApJ...699..626H}
---. 2009, \apj, 699, 626, \dodoi{10.1088/0004-637X/699/1/626}

\bibitem[{{Kaaret} {et~al.}(2017){Kaaret}, {Feng}, \&
  {Roberts}}]{2017ARA&A..55..303K}
{Kaaret}, P., {Feng}, H., \& {Roberts}, T.~P. 2017, \araa, 55, 303,
  \dodoi{10.1146/annurev-astro-091916-055259}

\bibitem[{{Kennicutt} \& {Evans}(2012)}]{2012ARA&A..50..531K}
{Kennicutt}, R.~C., \& {Evans}, N.~J. 2012, \araa, 50, 531,
  \dodoi{10.1146/annurev-astro-081811-125610}

\bibitem[{{Kimbro} {et~al.}(2021){Kimbro}, {Reines}, {Molina}, {Deller}, \&
  {Stern}}]{2021ApJ...912...89K}
{Kimbro}, E., {Reines}, A.~E., {Molina}, M., {Deller}, A.~T., \& {Stern}, D.
  2021, \apj, 912, 89, \dodoi{10.3847/1538-4357/abec6a}

\bibitem[{{Latimer} {et~al.}(2021{\natexlab{a}}){Latimer}, {Reines}, {Bogdan},
  \& {Kraft}}]{2021ApJ...922L..40L}
{Latimer}, L.~J., {Reines}, A.~E., {Bogdan}, A., \& {Kraft}, R.
  2021{\natexlab{a}}, \apjl, 922, L40, \dodoi{10.3847/2041-8213/ac3af6}

\bibitem[{{Latimer} {et~al.}(2021{\natexlab{b}}){Latimer}, {Reines},
  {Hainline}, {Greene}, \& {Stern}}]{2021ApJ...914..133L}
{Latimer}, L.~J., {Reines}, A.~E., {Hainline}, K.~N., {Greene}, J.~E., \&
  {Stern}, D. 2021{\natexlab{b}}, \apj, 914, 133,
  \dodoi{10.3847/1538-4357/abfe0c}

\bibitem[{{Lemons} {et~al.}(2015){Lemons}, {Reines}, {Plotkin}, {Gallo}, \&
  {Greene}}]{2015ApJ...805...12L}
{Lemons}, S.~M., {Reines}, A.~E., {Plotkin}, R.~M., {Gallo}, E., \& {Greene},
  J.~E. 2015, \apj, 805, 12, \dodoi{10.1088/0004-637X/805/1/12}

\bibitem[{{Mezcua} {et~al.}(2018){Mezcua}, {Civano}, {Marchesi}, {Suh},
  {Fabbiano}, \& {Volonteri}}]{2018MNRAS.478.2576M}
{Mezcua}, M., {Civano}, F., {Marchesi}, S., {et~al.} 2018, \mnras, 478, 2576,
  \dodoi{10.1093/mnras/sty1163}

\bibitem[{{Moretti} {et~al.}(2003){Moretti}, {Campana}, {Lazzati}, \&
  {Tagliaferri}}]{2003ApJ...588..696M}
{Moretti}, A., {Campana}, S., {Lazzati}, D., \& {Tagliaferri}, G. 2003, \apj,
  588, 696, \dodoi{10.1086/374335}

\bibitem[{{Park} {et~al.}(2006){Park}, {Kashyap}, {Siemiginowska}, {van Dyk},
  {Zezas}, {Heinke}, \& {Wargelin}}]{2006ApJ...652..610P}
{Park}, T., {Kashyap}, V.~L., {Siemiginowska}, A., {et~al.} 2006, \apj, 652,
  610, \dodoi{10.1086/507406}

\bibitem[{{Predehl} {et~al.}(2021){Predehl}, {Andritschke}, {Arefiev},
  {Babyshkin}, {Batanov}, {Becker}, {B{\"o}hringer}, {Bogomolov}, {Boller},
  {Borm}, {Bornemann}, {Br{\"a}uninger}, {Br{\"u}ggen}, {Brunner}, {Brusa},
  {Bulbul}, {Buntov}, {Burwitz}, {Burkert}, {Clerc}, {Churazov}, {Coutinho},
  {Dauser}, {Dennerl}, {Doroshenko}, {Eder}, {Emberger}, {Eraerds},
  {Finoguenov}, {Freyberg}, {Friedrich}, {Friedrich}, {F{\"u}rmetz},
  {Georgakakis}, {Gilfanov}, {Granato}, {Grossberger}, {Gueguen}, {Gureev},
  {Haberl}, {H{\"a}lker}, {Hartner}, {Hasinger}, {Huber}, {Ji}, {Kienlin},
  {Kink}, {Korotkov}, {Kreykenbohm}, {Lamer}, {Lomakin}, {Lapshov}, {Liu},
  {Maitra}, {Meidinger}, {Menz}, {Merloni}, {Mernik}, {Mican}, {Mohr},
  {M{\"u}ller}, {Nandra}, {Nazarov}, {Pacaud}, {Pavlinsky}, {Perinati},
  {Pfeffermann}, {Pietschner}, {Ramos-Ceja}, {Rau}, {Reiffers}, {Reiprich},
  {Robrade}, {Salvato}, {Sanders}, {Santangelo}, {Sasaki}, {Scheuerle},
  {Schmid}, {Schmitt}, {Schwope}, {Shirshakov}, {Steinmetz}, {Stewart},
  {Str{\"u}der}, {Sunyaev}, {Tenzer}, {Tiedemann}, {Tr{\"u}mper}, {Voron},
  {Weber}, {Wilms}, \& {Yaroshenko}}]{2021A&A...647A...1P}
{Predehl}, P., {Andritschke}, R., {Arefiev}, V., {et~al.} 2021, \aap, 647, A1,
  \dodoi{10.1051/0004-6361/202039313}

\bibitem[{{Reines}(2022)}]{2022NatAs...6...26R}
{Reines}, A.~E. 2022, Nature Astronomy, 6, 26,
  \dodoi{10.1038/s41550-021-01556-0}

\bibitem[{{Reines} {et~al.}(2020){Reines}, {Condon}, {Darling}, \&
  {Greene}}]{2020ApJ...888...36R}
{Reines}, A.~E., {Condon}, J.~J., {Darling}, J., \& {Greene}, J.~E. 2020, \apj,
  888, 36, \dodoi{10.3847/1538-4357/ab4999}

\bibitem[{{Reines} {et~al.}(2013){Reines}, {Greene}, \&
  {Geha}}]{2013ApJ...775..116R}
{Reines}, A.~E., {Greene}, J.~E., \& {Geha}, M. 2013, \apj, 775, 116,
  \dodoi{10.1088/0004-637X/775/2/116}

\bibitem[{{Reines} {et~al.}(2014){Reines}, {Plotkin}, {Russell}, {Mezcua},
  {Condon}, {Sivakoff}, \& {Johnson}}]{2014ApJ...787L..30R}
{Reines}, A.~E., {Plotkin}, R.~M., {Russell}, T.~D., {et~al.} 2014, \apjl, 787,
  L30, \dodoi{10.1088/2041-8205/787/2/L30}

\bibitem[{{Reines} {et~al.}(2016){Reines}, {Reynolds}, {Miller}, {Sivakoff},
  {Greene}, {Hickox}, \& {Johnson}}]{2016ApJ...830L..35R}
{Reines}, A.~E., {Reynolds}, M.~T., {Miller}, J.~M., {et~al.} 2016, \apjl, 830,
  L35, \dodoi{10.3847/2041-8205/830/2/L35}

\bibitem[{{Reines} \& {Volonteri}(2015)}]{2015ApJ...813...82R}
{Reines}, A.~E., \& {Volonteri}, M. 2015, \apj, 813, 82,
  \dodoi{10.1088/0004-637X/813/2/82}

\bibitem[{{Sacchi} {et~al.}(2024){Sacchi}, {Bogdan}, {Chadayammuri}, \&
  {Ricarte}}]{2024arXiv240601707S}
{Sacchi}, A., {Bogdan}, A., {Chadayammuri}, U., \& {Ricarte}, A. 2024, arXiv
  e-prints, arXiv:2406.01707, \dodoi{10.48550/arXiv.2406.01707}

\bibitem[{{She} {et~al.}(2017){She}, {Ho}, \& {Feng}}]{2017ApJ...835..223S}
{She}, R., {Ho}, L.~C., \& {Feng}, H. 2017, \apj, 835, 223,
  \dodoi{10.3847/1538-4357/835/2/223}

\bibitem[{{Silk}(2017)}]{2017ApJ...839L..13S}
{Silk}, J. 2017, \apjl, 839, L13, \dodoi{10.3847/2041-8213/aa67da}

\bibitem[{{Swartz} {et~al.}(2004){Swartz}, {Ghosh}, {Tennant}, \&
  {Wu}}]{2004ApJS..154..519S}
{Swartz}, D.~A., {Ghosh}, K.~K., {Tennant}, A.~F., \& {Wu}, K. 2004, \apjs,
  154, 519, \dodoi{10.1086/422842}

\bibitem[{{Swartz} {et~al.}(2008){Swartz}, {Soria}, \&
  {Tennant}}]{2008ApJ...684..282S}
{Swartz}, D.~A., {Soria}, R., \& {Tennant}, A.~F. 2008, \apj, 684, 282,
  \dodoi{10.1086/587776}

\bibitem[{{Thygesen} {et~al.}(2023){Thygesen}, {Plotkin}, {Soria}, {Reines},
  {Greene}, {Anderson}, {Baldassare}, {Owens}, {Urquhart}, {Gallo},
  {Miller-Jones}, {Paul}, \& {Rollings}}]{2023MNRAS.519.5848T}
{Thygesen}, E., {Plotkin}, R.~M., {Soria}, R., {et~al.} 2023, \mnras, 519,
  5848, \dodoi{10.1093/mnras/stad002}

\bibitem[{{Volonteri}(2010)}]{2010A&ARv..18..279V}
{Volonteri}, M. 2010, \aapr, 18, 279, \dodoi{10.1007/s00159-010-0029-x}

\end{thebibliography}
\end{document}